\documentclass[aps,prl,twocolumn,superscriptaddress,showpacs,floatfix]{revtex4-1}
\usepackage{graphicx}
\usepackage{lineno}
\usepackage{color}
\usepackage{upgreek}

\begin{document}

\title{Ground state selection under pressure in the quantum pyrochlore magnet Yb$_2$Ti$_2$O$_7$}

\author{E.~Kermarrec}
 \affiliation{Laboratoire de Physique des Solides,  CNRS, Univ. Paris-Sud, Universit\'e  Paris-Saclay, 91405 Orsay Cedex, France}
 \affiliation{Department of Physics and Astronomy, McMaster University, Hamilton, Ontario, L8S 4M1, Canada}
 \affiliation{Laboratoire National des Champs Magn\'etiques Intenses, CNRS, BP 166-38042 Grenoble, France} 

\author{J.~Gaudet}
\affiliation{Department of Physics and Astronomy, McMaster University, Hamilton, Ontario, L8S 4M1, Canada}

\author{K.~Fritsch}
\affiliation{Helmholtz-Zentrum Berlin f\"ur  Materialien und Energie, Hahn-Meitner-Platz 1, 14109 Berlin, Germany}

\author{R.~Khasanov}
\affiliation{Laboratory for Muon Spin Spectroscopy, Paul Scherrer Institut, CH-5232 Villigen PSI, Switzerland}

\author{Z.~Guguchia}
\affiliation{Laboratory for Muon Spin Spectroscopy, Paul Scherrer Institut, CH-5232 Villigen PSI, Switzerland}

\author{C.~Ritter}
\affiliation{Institut Laue Langevin, BP 156, 38042 Grenoble, France}

\author{K.~A.~Ross}
\affiliation{Department of Physics, Colorado State University, Fort Collins, Colorado 80523-1875, USA}

\author{H.~A.~Dabkowska}
\affiliation{Brockhouse Institute for Materials Research, Hamilton, Ontario L8S 4M1, Canada}

\author{B.~D.~Gaulin}
\affiliation{Department of Physics and Astronomy, McMaster University, Hamilton, Ontario, L8S 4M1, Canada}
\affiliation{Brockhouse Institute for Materials Research, Hamilton, Ontario L8S 4M1, Canada}
\affiliation{Canadian Institute for Advanced Research, 180 Dundas Street West, Toronto, Ontario M5G 1Z8, Canada}

%\begin{affiliations}
%\begin{footnotesize}
% \item Laboratoire de Physique des Solides,  CNRS, Univ. Paris-Sud, Universit\'e  Paris-Saclay, 91405 Orsay Cedex, France
% \item Department of Physics and Astronomy, McMaster University, Hamilton, Ontario, L8S 4M1, Canada
% \item Laboratoire National des Champs Magn\'etiques Intenses, CNRS, BP 166-38042 Grenoble, France 
% \item Helmholtz-Zentrum Berlin f\"ur  Materialien und Energie, Hahn-Meitner-Platz 1, 14109 Berlin, Germany 
% \item Laboratory for Muon Spin Spectroscopy, Paul Scherrer Institut, CH-5232 Villigen PSI, Switzerland
% \item Institut Laue Langevin, BP 156, 38042 Grenoble, France
% \item Department of Physics, Colorado State University, Fort Collins, Colorado 80523-1875, USA 
% \item Brockhouse Institute for Materials Research, Hamilton, Ontario, L8S 4M1, Canada 
% \item Canadian Institute for Advanced Research, 180 Dundas St.\ W., Toronto, Ontario, M5G 1Z8, Canada
% \item $^{\star}$email: edwin.kermarrec@u-psud.fr, bruce.gaulin@gmail.com
%\end{footnotesize} 
%\end{affiliations}

%\linenumbers

\begin{abstract}
%For Nature, the abstract is really an introductory paragraph set
%in bold type.  This paragraph must be ``fully referenced'' and
%less than 180 words for Letters.  This is the thing that is
%supposed to be aimed at people from other disciplines and is
%arguably the most important part to getting your paper past the
%editors.  End this paragraph with a sentence like ``Here we
%show...'' or something similar.
%
A quantum spin liquid is a novel state of matter characterized by quantum entanglement and the absence of any broken symmetry. In condensed matter, the frustrated rare-earth pyrochlore magnets Ho$_2$Ti$_2$O$_7$ and Dy$_2$Ti$_2$O$_7$, so-called spin ices, exhibit a classical spin liquid state with fractionalized thermal excitations (magnetic monopoles).  Evidence for a quantum spin ice, in which the magnetic monopoles become long range entangled and an emergent quantum electrodynamics arises, seems within reach. The magnetic properties of the quantum spin ice candidate Yb$_2$Ti$_2$O$_7$ have eluded a global understanding and even the presence or absence of static magnetic order at low temperatures is controversial. Here we show that sensitivity to pressure is the missing key to the low temperature behaviour of Yb$_2$Ti$_2$O$_7$. By combining neutron diffraction and muon spin relaxation on a stoichiometric sample under pressure, we evidence a magnetic transition from a disordered, non-magnetic, ground state to a splayed ferromagnetic ground state.  
\end{abstract}


\maketitle

%Then the body of the main text appears after the intro paragraph.
%Figure environments can be left in place in the document.
%\verb|\includegraphics| commands are ignored since Nature wants
%the figures sent as separate files and the captions are
%automatically moved to the end of the document (they are printed
%out with the \verb|\end{document}| command. However, tables must
%be manually moved to the end of the document, after the addendum.
%
%Citation of Einstein's paper \cite{Einstein}.

\section*{Introduction}

\begin{figure*}
\includegraphics[width=0.8\paperwidth]{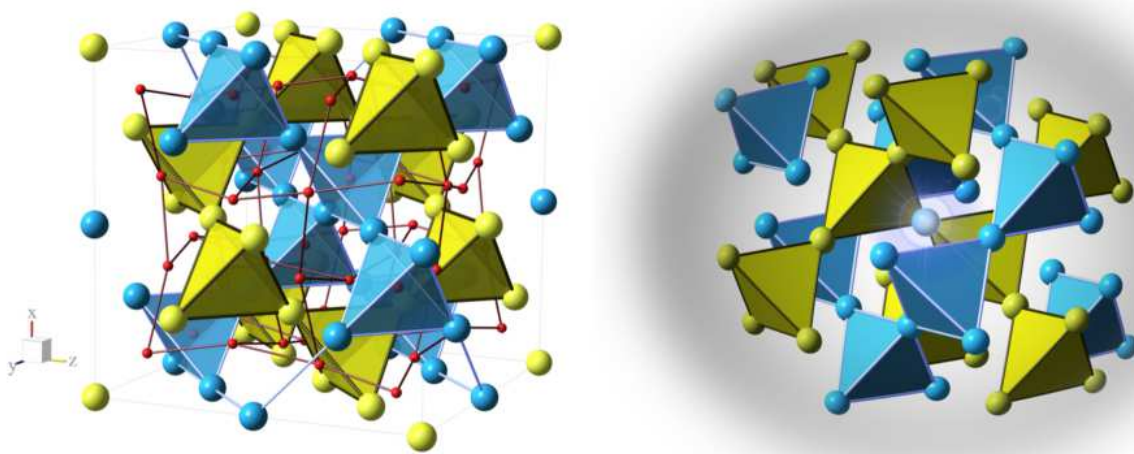} % this command will be ignored
\caption{\textbf{Pyrochlore structure of Yb$_{2+x}$Ti$_{2-x}$O$_{7+\delta}$.} Excess Yb$^{3+}$ ion can occupy a Ti$^{4+}$ site and create a local defect (referred to as ``stuffing''). Left: Representation of the ideal pyrochlore structure of Yb$_{2}$Ti$_{2}$O$_{7}$, with Yb in blue, Ti in yellow and O in red. Yb and Ti form corner sharing tetrahedra lattices. Right: Schematic representation of the structurally distorted environment of a defect.}
\label{fig_crystallo}
\end{figure*}

The pyrochlore lattice, comprised of corner-sharing tetrahedra, is the archetype of magnetic frustration in three dimensions\cite{IFM} (Fig.\ref{fig_crystallo}). Since its early study by Anderson in 1956,\cite{Anderson1956} frustrated spin Hamiltonians on the pyrocholore lattice have provided a seemingly-inexhaustible source for the study of fundamental physics.\cite{Bramwell2001,Fennell2009}
In particular,  spin liquid ground states have been predicted for such a lattice decorated with Heisenberg\cite{Moessner1998,Canals1998} or XXZ\cite{Hermele2004} spins. More recently, pyrochlore magnets have been put forward as realistic vehicles for the realization of a quantum spin ice (QSI) state, using the generic $S=\frac{1}{2}$ nearest-neighbour anisotropic exchange Hamiltonian\cite{Gingras2014,Savary2012,Benton2012}.
Yb$_2$Ti$_2$O$_7$ is a promising quantum spin ice candidate as it possesses both an (effective) $S=\frac{1}{2}$ spin, thanks to the well isolated crystal field Kramers doublet ground state appropriate to Yb$^{3+}$,\cite{Gaudet2015} and strong quantum fluctuations brought by anisotropic exchange interactions and an XY $g$-tensor.\cite{RossPRX2012} 
Several studies have focused on the nature of the ground state in Yb$_2$Ti$_2$O$_7$, yet a consensus has been elusive to date.\cite{Hodges2002,Yasui2003,Gardner2004,Chang2012,D'Ortenzio2013,Chang2014} Early neutron scattering experiments ruled out the presence of conventional static order down to 90mK in a polycrystalline sample\cite{Gardner2004}, whereas other single crystal studies concluded the ground state was ferromagnetic \cite{Chang2012, Yasui2003}. The results of local probes are even more puzzling. Muon spin relaxation ($\upmu$SR) measurements evidenced the presence of true static moments on the muon timescale, through the observation of both a drop of asymmetry and a decoupling of the muon spins in longitudinal applied fields\cite{Chang2014}, along with a drastic slowing down of the fluctuation rate below $T_{\rm c}$ for certain samples\cite{Hodges2002}. In contrast, $\upmu$SR studies by D'Ortenzio \textit{et al.}\cite{D'Ortenzio2013} found a non-magnetic, fluctuating ground state, in both stoichiometric polycrystalline and single crystal samples, \textit{despite} the presence of pronounced specific heat anomalies at $T_{\rm c} = 265$~mK and $T_{\rm c} = 185$~mK, respectively.
It is clear that local defects, either oxygen vacancies\cite{Sala2014} or excess magnetic ions\cite{RossPRB2012} (referred to as stuffing),  vary significantly between polycrystalline powders and single crystals, and are likely responsible for such sample dependencies.
Here, by applying hydrostatic pressure to well-characterized Yb$_{2+x}$Ti$_{2-x}$O$_{7+\delta}$ samples, with $x=0$ and $x=0.046$\cite{RossPRB2012}, we observe a magnetic transition in the stoichiometric, $x=0$ sample from a disordered ground state into a splayed ferromagnetic ground state. This result sheds light on the origin of the sample dependence in the ground state selection for Yb$_2$Ti$_2$O$_7$ and is consistent with the recent theoretical proposal that Yb$_2$Ti$_2$O$_7$ lies close to a phase boundary in the generic QSI Hamiltonian phase diagram\cite{Jaubert2015}. 

\section*{Results}

\subsection{Muon spin relaxation}
\begin{figure*}
\includegraphics[width=0.8\paperwidth]{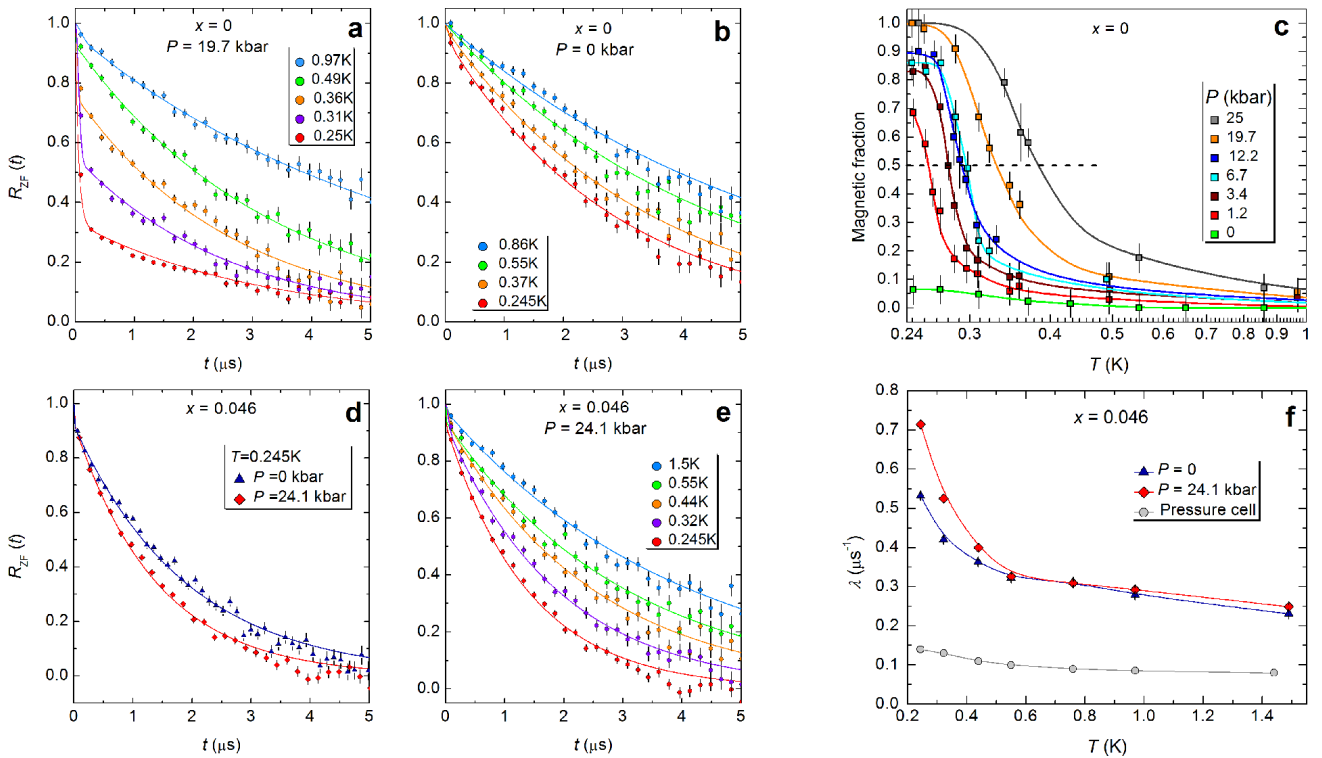} % this command will be ignored
\caption{\textbf{Temperature evolution of the $\upmu$SR relaxation in Yb$_{2+x}$Ti$_{2-x}$O$_{7+\delta}$ under applied pressure}.  Panels a,b and c (d, e and f) refers to the $x=0$ ($x=0.046$) sample. (a) The drastic increase of relaxation observed upon decreasing temperature in the $x=0$ sample indicates a spin freezing under an applied pressure $P = 19.7$ kbar, which is absent under zero pressure (b) and for the $x=0.046$ sample (d). (c) The temperature evolution of the magnetic fraction is reported for various pressures. The black horizontal dashed line represents a volume magnetic fraction of 50\%, used as a criterion to define $T_{\rm c}$. (e,f) For the $x=0.046$ sample, only a moderate increase of the spin dynamics is observed under applied pressure.  The error bars of the $\upmu$SR relaxation data are of statistical origin and correspond to the square root of the total number of detected positrons resulting from muon decays.  Error bars of the relaxation rate $\lambda$ represent standard deviation of the fit parameters. Error bars of the magnetic fraction represent standard deviation of the fit parameters, with a minimal value of 0.05 corresponding to the typical error on the total asymmetry for $\upmu$SR under pressure.}
\label{fig_musr}
\end{figure*}

$\upmu$SR measurements under hydrostatic pressures as high as 25 kbar, and at temperatures as low as 0.245 K were performed on Yb$_{2+x}$Ti$_{2-x}$O$_{7+\delta}$ samples, with $x=0$ and $x=0.046$ at the GPD beamline of PSI. The muons are implanted inside the bulk of the material, and act as local magnetic probes. The signal coming from the muons that stop inside the pressure cell was measured separately and subtracted (see Supplementary Fig. 1 and Supplementary Fig. 2) from the overall signal.

\begin{figure}
\includegraphics[width=\columnwidth]{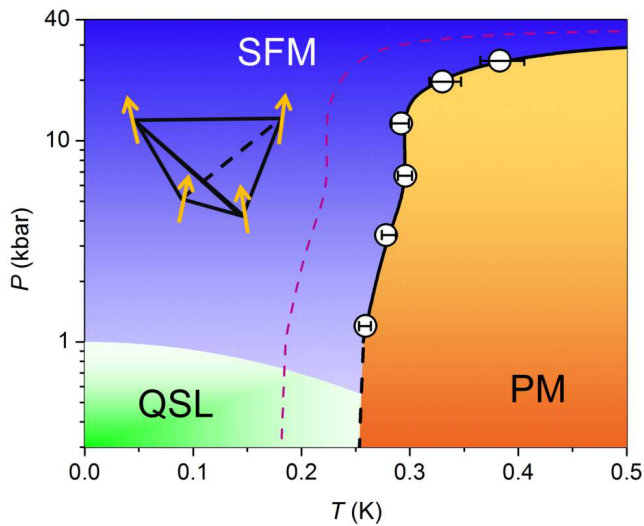} % this command will be ignored
\caption{\textbf{Pressure-temperature phase diagram of Yb$_{2+x}$Ti$_{2-x}$O$_{7+\delta}$}. The vertical axis displays the pressure $P$ in logarithmic scale and the horizontal axis the temperature $T$. Empty black circles define the transition line between the collective paramagnetic (PM, orange) and the splayed ferromagnetic (SFM, blue) regions relative to the $x=0$ sample. The transition temperature is defined such that for $T \leq T_{\rm c}$ 50\% of the magnetic moments are frozen (see Fig.\ref{fig_musr} c). Error bars allows $T_{\rm c}$ to be defined between 40\% and 60\% of the magnetic fraction. The green region  highlights the presence of a disordered, non-magnetic, phase (QSL) found at $P=0$. Black thick line is a guide to the eye. Dashed purple line is the hypothetical transition line for $x=0.046$.}
\label{fig_diag}
\end{figure}

Fig.\ref{fig_musr}a shows the temperature dependence of the muon spin relaxation for the stoichiometric, $x=0$ sample in zero field, $R_{\rm zf}(t)$, as a function of time $t$ and under an applied pressure $P=19.7$~kbar. Well above $T_{\rm c} = 0.265$ K, at $T \geq 0.97$~K, the majority of the Yb$^{3+}$ magnetic moments are paramagnetic and in a fast fluctuating regime, and display single-exponential relaxation. For $T \leq 0.5$~K, we observe the development of a small magnetic fraction $f$ of the Yb$^{3+}$ moments, which grows non-linearly as the temperature decreases. The absence of oscillations at short time is indicative of a highly disordered magnetic state. The zero-field relaxation is well described by a Gaussian distribution of static internal fields with standard deviation $\Delta$ (see Supplementary Note 1), and the following phenomenological function:
\begin{equation}
{R_{\rm zf}}(t) = f \left( \frac{2}{3} e^{-\Delta^2 t^2 /2} + \frac{1}{3} e^{-\lambda t} \right) + (1-f)e^{-\lambda t} \\
\label{Pzf}
\end{equation} 
In a purely static scenario, the second term (1/3-tail) should be constant. Here, a fluctuating component is nonetheless observed and we modelled this using a relaxation rate $\lambda$. The third term accounts for the paramagnetic component, and assumes the same relaxation rate $\lambda$, for simplicity. The unconventional shape of the zero-field longitudinal relaxation was discussed in detail in Ref\cite{Hodges2002,Yaouanc2013}.    
In contrast, the evolution of the relaxation in temperature of the $x=0$ sample under zero applied pressure, in Fig.\ref{fig_musr}b, shows little or no magnetic fraction ($f \simeq 6$\%) at any temperature above our base $T=0.245$~K, in agreement with D'Ortenzio et al.'s previously reported $\upmu$SR studies. Using equation (\ref{Pzf}) we extract the magnetic fraction for each pressure and temperature, and collect the results in Fig.\ref{fig_musr}c. The development of the magnetic fraction with temperature is clearly pressure dependent, and turns on strongly at low temperatures, below $T_{\rm c} = 0.265$ K, for our minimum pressure of 1.2 kbar. For each pressure, one can define a critical temperature $T_{\rm c}$, such that for $T \leq T_{\rm c}$, 50\% of the magnetic moments are frozen. The corresponding $P-T$ phase diagram is shown in Fig.\ref{fig_diag}. Clearly, the phase transition extrapolated from finite pressure measurements to zero pressure, agrees well with the sharp $C_{\rm p}$ anomaly at $T_{\rm c} = 0.265$ K, appropriate to the $x=0$ sample. However the zero-pressure state for the $x=0$ sample at 0.245 K, below $T_{\rm c}$, is disordered, indicating that the ground state of the stoichiometric, $x=0$ sample, is a spin liquid. 

%The evolution of the magnetic fraction with temperature at $P \sim 12$~kbar is in very good agreement with the order parameter previously determined with neutron diffraction (Fig. 2b).  This correlation strongly suggests that the continuous increase of the neutron elastic intensity upon cooling is due to the growth of magnetically frozen parts in the sample at the expense of paramagnetic ones.
% rather than to a conventional second-order transition. This is therefore compatible with a first-order character of the transition.  
 
We now turn to the $x=0.046$ sample. The zero-field relaxation at $T=0.245$~K under zero and an applied pressure $P=24.1$~kbar are shown in Fig.\ref{fig_musr}d. Strikingly, no frozen magnetic fraction is observed upon the application of a pressure as high as $P=24.1$~kbar. Instead, we observe an increase of the relaxation for this $x=0.046$ sample, demonstrating its sensitivity to pressure. The temperature dependence of the relaxation is reported in Fig.\ref{fig_musr}e and f. One can speculate that a transition to a fully ordered state, as it is observed for the $x=0$ sample, would require higher pressures or lower temperatures, consistent with the lower $T_{\rm c} =0. 185$~K of the $x=0.046$ sample . 

$\upmu$SR studies on other samples have reported a drastic slowing down of spin fluctuations\cite{Hodges2002}, or static order\cite{Chang2014}, under \textit{zero} applied pressure for temperatures below 0.25~K. In the light of our results, even relatively low (applied or chemical) pressure can destroy the disordered spin liquid state and induce magnetic order.  A low level of defects in the different samples is a natural explanation to the contradictory $\upmu$SR results. Such disorder, at the $\sim$ 2 \% level, is difficult to characterize, but it is largely absent in polycrystalline samples, synthesized at lower temperatures by solid state methods.

%More sophisticated crystallographic techniques such as neutron PDF or (additional\cite{Chang2012}) EXAFS would appear necessary to understand the local chemical pressure effect of a 2\% level of disorder.    

\subsection{Neutron diffraction}

\begin{figure*}
\includegraphics[width=0.75\paperwidth]{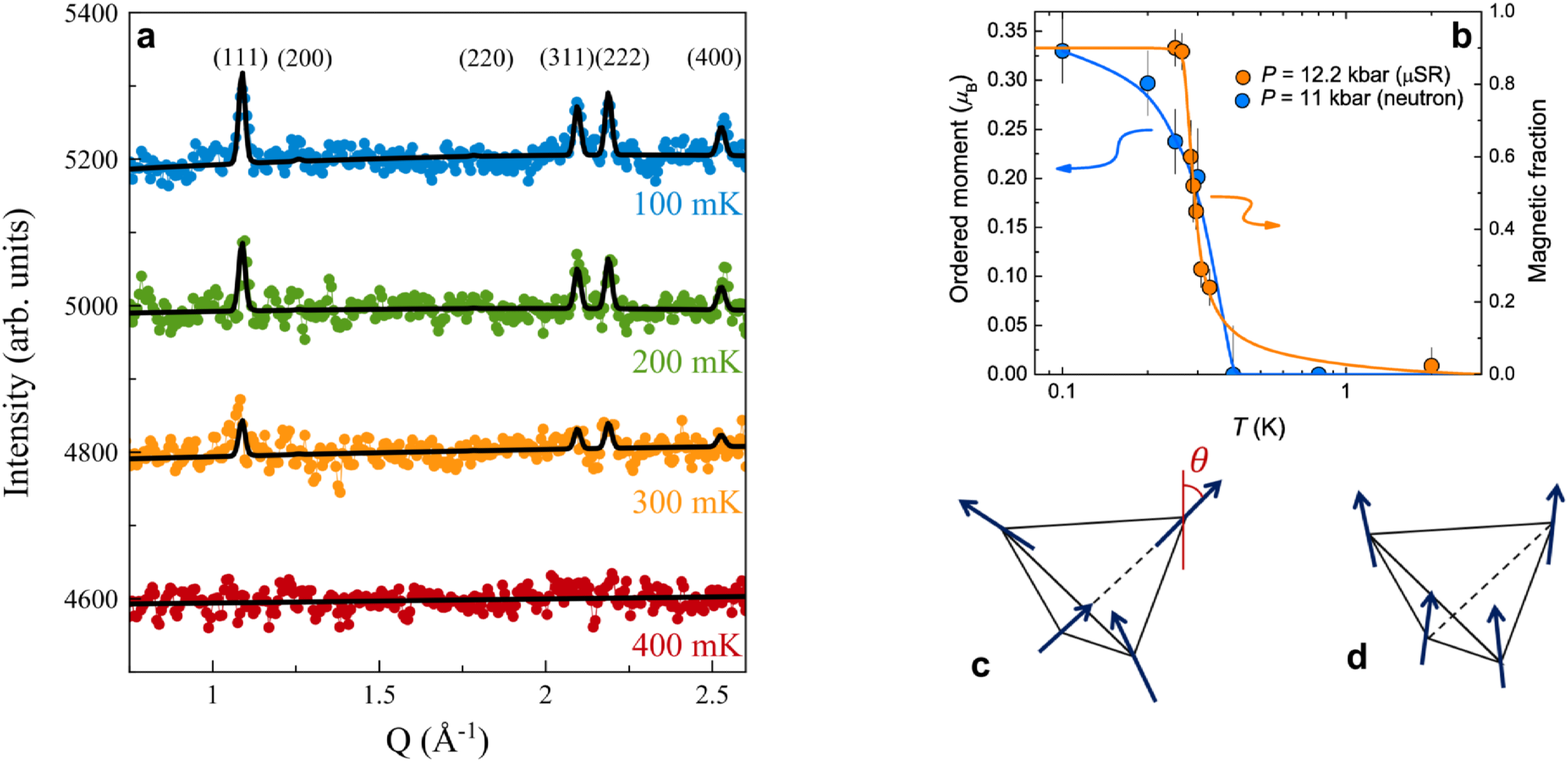} % this command will be ignored
\caption{\textbf{Neutron diffraction measurements of Yb$_{2}$Ti$_{2}$O$_{7}$ under applied pressure.}
a) Diffraction datasets from 400 to 100 mK after the 800 mK dataset has been subtracted. Error bars are not shown for clarity.
b) Ordered moment vs. temperature determined by neutron diffraction for $P=11(2)$ kbar (blue, left axis) and magnetic fraction determined by $\upmu$SR for $P=12.2$ kbar (orange, right axis). Error bars of the ordered moment represent standard deviation of the refinement.
Schematic spin structure of the ice-like splayed ferromagnet with $\theta = 14^{\circ}$ for $P=0$ (c) and  $\theta = 5^{\circ}$ for $P=11$~kbar (d), where $\theta$ is the splay angle between the [001] direction and the magnetic moment, tilted towards the local [111] directions of the tetrahedron. 
}
\label{fig_neutron}
\end{figure*}

Armed with the knowledge of the $P-T$ phase diagram in Fig.\ref{fig_diag}, we sought to determine the nature of the pressure-induced magnetic order in Yb$_{2+x}$Ti$_{2-x}$O$_{7+\delta}$ samples, with $x=0$, by performing neutron diffraction on the stoichiometric powder sample at the D20 high-flux diffractometer of the ILL.  The detection of small magnetic moments under pressure using neutron diffraction is challenging due to the significant background signal of the pressure cell itself. Fig.\ref{fig_neutron}a shows the neutron diffraction data for the maximum hydrostatic pressure of the cell, $P=11(2)$ kbar, and  temperatures from 400 to 100mK, from which a background measured at 800mK was subtracted. We clearly observe the development of magnetic Bragg intensities at the (111), (311), (222) and (004) positions upon cooling below 400mK. This is firm evidence for the existence of long range magnetic order in Yb$_{2+x}$Ti$_{2-x}$O$_{7+\delta}$ samples, with $x=0$, under an applied pressure $P=11(2)$~kbar. The refinement of the neutron diffraction data gives us the temperature dependence of the ordered moment, shown in Fig.\ref{fig_neutron}b.
The contrast with previous experiments under zero pressure is striking. First, the saturated moment $\mu = 0.33(5)$~$\mu_{\rm B}$ is much smaller  than that $\mu \sim 1$~$\mu_{\rm B}$ reported previously for different Yb$_2$Ti$_2$O$_7$ samples\cite{Chang2012}, although similar to the ordered moment in the $\Gamma_5$ ordered state of Yb$_2$Ge$_2$O$_7$. \cite{new_Hallas_PRB} Second, the ordered moment vanishes cleanly above $T_{\rm c} \sim 0.4$~K, with no anomalous magnetic Bragg intensity well above $T_{\rm c}$ \cite{RossPRB2011,Gaudet2016}.
The previously reported  order parameter at $P=0$ of our $x=0$ polycrystalline sample is anomalous \cite{Gaudet2016}; it shows no change across $T_{\rm c}$ and only falls off at much higher temperatures.  Consistency with our $P=0$ $\upmu$SR results on the same sample requires that this Bragg-like scattering is dynamic on slow time scales.  That notwithstanding, the magnetic structure previously refined on the basis of a very high temperature ($\sim$ 8 K) background subtraction gave a splayed ice-like ferromagnetic structure\cite{Gaudet2016}, with the moments on a tetrahedron lying mainly in the [100] direction with a positive splay angle $\theta = 14(5)^{\circ}$, such that the moments tilt towards the local [111] direction (Fig.\ref{fig_neutron}c). The components perpendicular to the local [111] axis obey the 2-in/2-out ice rule on a single tetrahedron. A different type of splayed ferromagnet, with the perpendicular components satisfying the all-in/all-out structure, has also been reported recently\cite{Yaouanc2016}, in addition to a nearly collinear ferromagnet ($\theta \sim 0^{\circ}$)\cite{Chang2012}, for other samples.  The magnetic structure associated with the true Bragg scattering we refine here in the stoichiometric $x=0$ sample under $P=11(2)$~kbar is also a splayed ice-like structure, but with a much reduced splay angle $\theta = 5(4)^{\circ}$, such that it is close to a collinear [100] ferromagnet (Fig.\ref{fig_neutron}d).
%
%
%Under an applied pressure $P=11(2)$~kbar, the splayed angle is fairly reduced and the magnetic order becomes close to an almost perfect collinear ferromagnet with the moments pointing along [100], i.e with $\theta = 5$ (Fig.\ref{fig_neutron}d). 

\section*{Discussion}
%Our results are summarized in the pressure-temperature phase diagram of Fig.\ref{fig_diag}. 
%-Competition spin-ice ground-state
%-real Quenched disorder under pressure, only local minor effects 
%-disordered induced like TTO, new parameter for the study of frustrated pyrochlores.
%-CFM ??

%Conclusion

%2)pressure modifiy exchanges and not anisotropy

These results bring a fresh perspective on the long standing debate about the presence or absence of static magnetic order in the quantum pyrochlore magnet Yb$_{2+x}$Ti$_{2-x}$O$_{7+\delta}$. The acute sensitivity to local (through the Yb$^{3+}$ stuffing)  or applied pressure is surprising. However, a corollary of our new $P-T$ phase diagram is that non-stoichiometric samples with non-zero chemical pressure can easily display an ambient applied pressure phase transition to a splayed ferromagnetic state at $T_{\rm c}$.  
Yet, this interpretation remains challenged by the fact that our $x=0.046$ sample does not show evidence for magnetic order at ambient pressure, and by previous reported observations of a magnetic transition in polycrystalline, likely $x \sim 0$, sample even under zero pressure\cite{Hodges2002,Chang2014}. This may indicate that the non-magnetic low temperature region of the phase diagram is extremely narrow, existing only for a certain range of $x$, whose absolute values are still to be determined. This would actually be reminiscent of the recent findings on the Tb$_{2+x}$Ti$_{2-x}$O$_{7+\delta}$ pyrochlore magnet, that has been shown to display an ordered phase that is extremely sensitive to disorder, appearing only for $0 < x < 0.01$.\cite{Taniguchi2013,Kermarrec2015,Takatsu2016}

Furthermore, the present work illustrates the relevance of applying hydrostatic pressure to tune the magnetic properties of frustrated pyrochlore compounds, a path that was followed by pioneering work on the other spin liquid candidate Tb$_2$Ti$_2$O$_7$.\cite{Mirebeau2002}
%
%Under such relatively low applied pressures the local charge environment of Yb$^{3+}$ is unlikely to be affected, and therefore the $XY$ magnetic anisotropy must be preserved, along with the isolated Kramers doublet ground state. 
%
In case of Yb$_2$Ti$_2$O$_7$, we found that the pressure tunes the delicate balance between the anisotropic exchanges of the quantum spin ice Hamiltonian, and selects a splayed ferromagnetic ground state away from the degenerate antiferromagnetic ground states manifold. %\cite{Wong2013,Han2013}
%and selects a splayed ferromagnetic ground state away from the U(1) manifold phases.
This scenario confirms recent theoretical proposals that Yb$_{2}$Ti$_{2}$O$_{7}$ lies close to phase boundaries derived from the generic $S_{\rm eff} =\frac{1}{2}$ quantum spin ice Hamiltonian\cite{Jaubert2015}, and provides the missing key to understand its exotic magnetic properties. Particularly appealing is the prediction that accidental degeneracies in the vicinity of these phase boundaries can lead to the emergence of a quantum spin liquid.\cite{Han2013} This would offer a natural explanation for a non-magnetic, disordered state under zero pressure in stoichiometric Yb$_{2}$Ti$_{2}$O$_{7}$ and recent observations of a continuum of gapless quantum excitations\cite{Gaudet2016,Robert2015} at low temperatures.

After publication we became aware of the work of Arpino et al. that study in details the effect of off-stoichiometry in Yb$_{2+x}$Ti$_{2-x}$O$_{7+\delta}$ samples\cite{Arpino2017}.

%\section*{Another Section}
%
%Sections can only be used in Articles.  Contributions should be
%organized in the sequence: title, text, methods, references,
%Supplementary Information line (if any), acknowledgements,
%interest declaration, corresponding author line, tables, figure
%legends.
%
%Spelling must be British English (Oxford English Dictionary)
%
%In addition, a cover letter needs to be written with the
%following:
%\begin{enumerate}
% \item A 100 word or less summary indicating on scientific grounds
%why the paper should be considered for a wide-ranging journal like
%\textsl{Nature} instead of a more narrowly focussed journal.
% \item A 100 word or less summary aimed at a non-scientific audience,
%written at the level of a national newspaper.  It may be used for
%\textsl{Nature}'s press release or other general publicity.
% \item The cover letter should state clearly what is included as the
%submission, including number of figures, supporting manuscripts
%and any Supplementary Information (specifying number of items and
%format).
% \item The cover letter should also state the number of
%words of text in the paper; the number of figures and parts of
%figures (for example, 4 figures, comprising 16 separate panels in
%total); a rough estimate of the desired final size of figures in
%terms of number of pages; and a full current postal address,
%telephone and fax numbers, and current e-mail address.
%\end{enumerate}
%
%See \textsl{Nature}'s website
%(\texttt{http://www.nature.com/nature/submit/gta/index.html}) for
%complete submission guidelines.

%\begin{methods}
\section*{Methods}
%Put methods in here.  If you are going to subsection it, use
%\verb|\subsection| commands.  Methods section should be less than
%800 words and if it is less than 200 words, it can be incorporated
%into the main text.

%Here is a description of a specific method used.  Note that the
%subsection heading ends with a full stop (period) and that the
%command is \verb|\subsection{}| not \verb|\subsection*{}|.

%\subsection{Method subsection.}

\subsection{Sample preparation}
The Yb$_{2+x}$Ti$_{2-x}$O$_{7+\delta}$ samples with $x=0$ and $x=0.046$ were prepared at the Brockhouse Institute for Materials Research, McMaster University. The $x=0$ powder sample was obtained through conventional solid-state reaction between pressed powders of Yb$_2$O$_3$ and TiO$_2$ sintered at 1200$^{\circ}$C in air. The $x=0.046$ powder sample was obtained by crushing a single crystal grown by the floating zone method in 4 atm of O$_2$ with a growth rate of 5mm/h. More details on the details of the synthesis and the characterization can be found in Ref.\cite{RossPRB2011}. 

\subsection{Muon spin relaxation}
$\upmu$SR measurements were carried out at the GPD instrument of the Paul Scherrer Institut, Switzerland. About 1 g of each powder sample was mixed with $\sim 3$ mm$^3$ of a pressure medium (Daphne 7373 oil) and placed inside the sample channel of a double-wall pressure cell. Two different cells were used, labelled as (1) and (2) (see Supplementary Note 1), and are described in more details in Ref.\cite{Khasanov2016}. The muon momentum was adjusted in order to obtain an optimal fraction of the muons stopping in the sample, with optimal values found at $p=106$ and $p=107$ MeV/\textit{c}. The relaxation of both cells were measured without any sample down to 0.245 K. The applied pressure was determined by measuring the superconducting transition temperature of a small piece of pure indium inserted in the sample channel.\cite{Khasanov2016}

\subsection{Neutron diffraction}
The neutron diffraction experiment was conducted at the D20 beamline, a high intensity two axis diffractometer, at the Institut Laue-Langevin, using a neutron wavelength  $\lambda = 2.4$ \AA. A mass of 1.5 g of Yb$_{2}$Ti$_{2}$O$_{7}$ powder sample and a small amount of NaCl powder, which serves as a pressure calibration, were both mounted in a high pressure clamp cell and inserted in a  $^3$He-$^4$He dilution fridge. Fluorinert was used as a pressure transmitter. A minimum of 12 hours of data was collected at each temperature. The diffraction pattern obtained for $T=800$mK is shown in Supplementary Fig.3. Structural refinements for both NaCl and Yb$_{2}$Ti$_{2}$O$_{7}$ and magnetic refinements for Yb$_{2}$Ti$_{2}$O$_{7}$ have been performed using the Fullprof program suite\cite{Fullprof}.

\subsection{Data availability}
The datasets generated during and/or analysed during the current study are available from the corresponding author on reasonable request.

\subsection{Competing financial interests}
The authors declare no competing financial interests.

Work at McMaster University was supported by NSERC of Canada. This work is based on experiments performed at  S$\upmu$S, Paul Scherrer Institute, Villigen, Switzerland and at the Institut Laue-Langevin, Grenoble, France. This project has received funding from the European Union's Seventh Framework Programme for research, technological development and demonstration under the NMI3-II Grant number 283883. EK acknowledges useful discussions with P. Mendels and F. Bert. 

The authors declare that they have no competing financial interests.

EK, JG and BDG wrote the manuscript. EK, JG, KF and BDG performed the neutron diffraction experiment. EK and BDG performed the $\upmu$SR experiment. KAR and HAD synthezised and characterized the samples. CR designed and performed the neutron scattering experiment. ZG and RK designed and performed the $\upmu$SR experiment. All the co-authors discussed the results and improved the manuscript.   

Correspondence should be addressed to Dr. Edwin Kermarrec (edwin.kermarrec@u-psud.fr) and Dr. Bruce D. Gaulin~(bruce.gaulin@gmail.com).

\bibliographystyle{naturemag}
%\bibliography{sample}

\newpage

%%%%%FIGURES%%%%%%%%%%%%
%%%%%%%%%%%%%%%%%%%%%%%$

%\begin{figure}
%\includegraphics[width=\columnwidth]{Fig5.png} % this command will be ignored
%\caption{\textbf{Refinement of the diffraction pattern for T$=$800mK }.}
%\label{fig_refine}
%\end{figure}

%%
%% TABLES
%%
%% If there are any tables, put them here.
%%

%\begin{table}
%\centering
%\caption{This is a table with scientific results.}
%\medskip
%\begin{tabular}{ccccc}
%\hline
%1 & 2 & 3 & 4 & 5\\
%\hline
%aaa & bbb & ccc & ddd & eee\\
%aaaa & bbbb & cccc & dddd & eeee\\
%aaaaa & bbbbb & ccccc & ddddd & eeeee\\
%aaaaaa & bbbbbb & cccccc & dddddd & eeeeee\\
%1.000 & 2.000 & 3.000 & 4.000 & 5.000\\
%\hline
%\end{tabular}
%\end{table}

\end{document}